\documentclass[aps,superscriptaddress,twocolumn]{revtex4}
\usepackage{graphics}
\usepackage{epsfig}
\usepackage{color}
\usepackage{stackrel}
\usepackage{amsfonts}
\usepackage{wrapfig}
\usepackage{pstricks}
\usepackage{pst-node}
\usepackage{bm}
\usepackage{dcolumn}
\usepackage{multirow}
\usepackage{epstopdf}
\usepackage{amssymb}
\usepackage{amsmath}
\usepackage{commath}
\usepackage{graphicx,bm}
\usepackage{verbatim}
\usepackage{booktabs}
\usepackage[para]{threeparttable}
\usepackage{etoolbox}
\usepackage{lipsum}
\usepackage{relsize}
\usepackage{braket}
\usepackage[caption=false]{subfig}
\usepackage{hyperref}
\usepackage[all]{hypcap}

\begin{document}

\title{The Coulomb interaction in monolayer transition-metal dichalcogenides}

\author{Dinh~Van~Tuan}
\affiliation{Department of Electrical and Computer Engineering, University of Rochester, Rochester, New York 14627, USA}

\author{Min~Yang}
\affiliation{Department of Electrical and Computer Engineering, University of Rochester, Rochester, New York 14627, USA}

\author{Hanan~Dery}
\altaffiliation{hanan.dery@rochester.edu}
\affiliation{Department of Electrical and Computer Engineering, University of Rochester, Rochester, New York 14627, USA}
\affiliation{Department of Physics and Astronomy, University of Rochester, Rochester, New York 14627, USA}

\begin{abstract}
Recently, the celebrated Rytova-Keldysh potential has been widely used to describe the Coulomb interaction of few-body complexes in monolayer transition-metal dichalcogenides. Using this potential to model charged excitons (trions), one finds a strong dependence of the binding energy on whether the monolayer is suspended in air, supported  on SiO$_2$, or encapsulated in hexagonal boron-nitride. However, empirical values of the trion binding energies show weak dependence on the monolayer configuration. This deficiency indicates that the description of the Coulomb potential is still lacking in this important class of materials. We address this problem and derive a new potential form, which takes into account the three atomic sheets that compose a monolayer of transition-metal dichalcogenides. The new potential self-consistently supports (i) the non-hydrogenic Rydberg series of neutral  excitons, and (ii) the weak dependence of the trion binding energy on the environment. Furthermore, we identify an important trion-lattice coupling due to the phonon cloud in the vicinity of charged complexes.  Neutral excitons in their ground state, on the other hand, have weaker coupling to the lattice due to the confluence of their charge neutrality and small Bohr radius. 
\end{abstract}
\pacs{71.45.Gm 71.10.-w  71.35.-y 78.55.-m}
\maketitle

\section{Introduction}
\begin{figure}
\includegraphics[width=8.5cm]{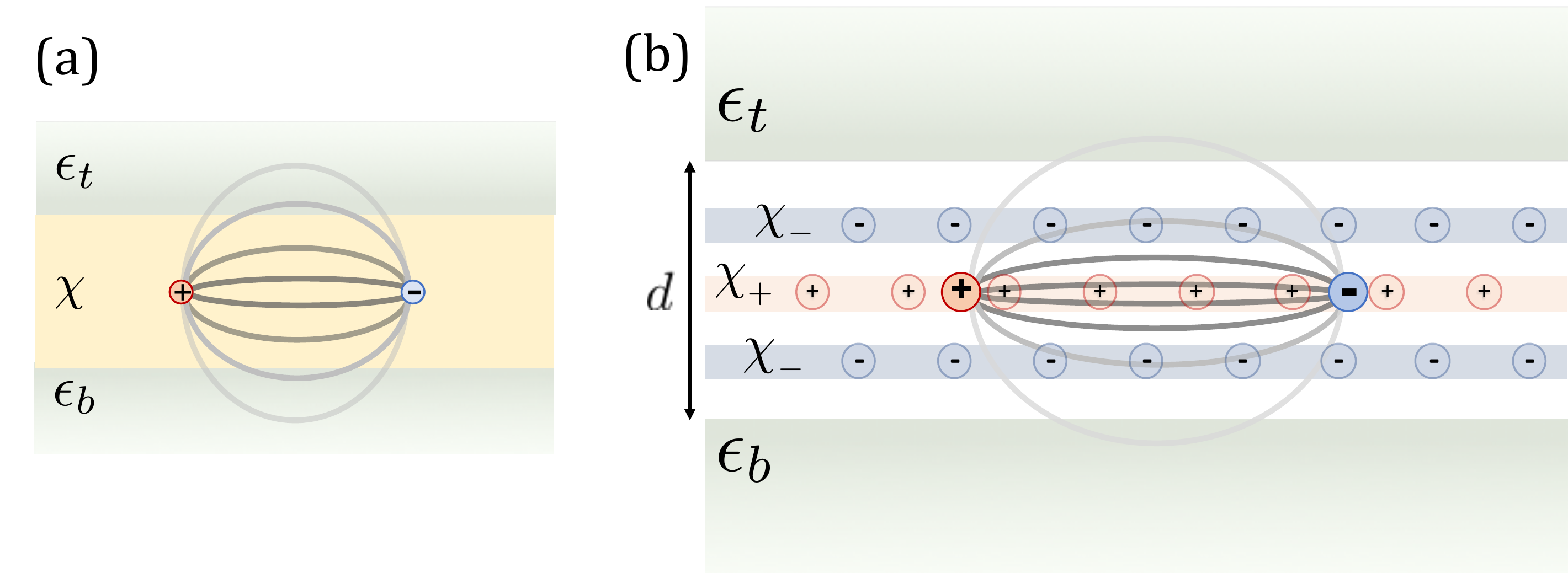}
 \caption{(a)  The dielectric environment when considering a uniform ML with polarizability $\chi$. Also shown are the field lines between opposite charges. (b) The ML thickness is $d$ and it is modeled as three atomic sheets with polarizabilities $\chi_+$ for the central one (Mo/W) and $\chi_-$ for the top and bottom ones (S/Se/Te, displaced by $\pm$d/4 from the center).  Screening from the chalcogen  sheets helps to confine the field lines in the ML, thereby reducing the dependence on the bottom and top materials whose dielectric constants are $\epsilon_b\,\&\,\epsilon_t$.} \label{fig:DielectricEnvironment}
\end{figure}

The discovery that monolayer transition-metal dichalcogenides (ML-TMDs) are two-dimensional (2D) direct band-gap semiconductors has sparked wide interest in their optical properties \cite{Splendiani_NanoLett10,Mak_PRL10,Korn_APL11,Wang_NatNano12,Xiao_PRL12,Britnell_Science13,Geim_Nature13,Zeng_NatNano12,Mak_NatNano12,Feng_NatComm12,Jones_NatNano13,Song_PRL13,Xu_NatPhys14}. It also resurfaced the problem of calculating exciton states in ultrathin semiconductor heterostructures by the use of the conventional Coulomb potential \cite{Rytova_MSU67,Keldysh_JETP79,SchmittRink_JL85,Cudazzo_PRB11,Berkelbach_PRB13,Chernikov_PRL14,Zhang_PRB14,Rosner_PRB15,Trolle_SR17,Meckbach_PRB18}, $e^2/\epsilon r$, where $e$ is the elementary charge, $\epsilon$ is an effective dielectric constant, and $r$ is the distance between the charged particles in the 2D plane. The conventional potential is a good description when the dielectric constants of the various layers have similar magnitudes. While this scenario holds in typical semiconductor quantum-well heterostructures such as Si/SiGe or GaAs/AlGaAs, it does not hold  when an ML-TMD or graphene is suspended in air, supported on low-dielectric materials or encapsulated between them.  The Rytova-Keldysh potential is a better description  in these cases \cite{Rytova_MSU67,Keldysh_JETP79,SchmittRink_JL85,Cudazzo_PRB11}, 
\begin{eqnarray}
V_{\text{RK}}(r)  = \frac{e_1e_2}{r_0} \frac{\pi}{2}  \left[ \mathbf{H}_0\left( \frac{\kappa r}{r_0}\right) - Y_0\left( \frac{\kappa r}{r_0}\right) \right ]\,,\,\,\,\,\, \label{eq:Keldysh}
\end{eqnarray}
where $e_1$ and $e_2$ are the charges of the interacting particles, located in the mid-plane of a thin semiconductor. The latter is embedded between top and bottom layers with dielectric constants $\epsilon_t$ and $\epsilon_b$, as shown in Fig.~\ref{fig:DielectricEnvironment}(a).  $\mathbf{H}_0$ and $Y_0$ are the zero-order Struve and Neumann special functions, $\kappa = (\epsilon_t+\epsilon_b)/2$, and $r_0$ is a measure of the dielectric screening length due to the polarizability of the 2D semiconductor \cite{Cudazzo_PRB11}.  

The Rytova-Keldysh potential has become a prevalent description of the Coulomb interaction in ML-TMDs after it was shown to support the non-hydrogenic Rydberg series of neutral excitons \cite{Berkelbach_PRB13,Chernikov_PRL14}.  However, in spite of its recent popularity \cite{Thilagam_JAP14,Wang_ADP14,Berghauser_PRB15,Zhang_NanoLett15,Ganchev_PRL15,Mayers_PRB15,Kylanpaa_PRB15,Velizhanin_PRB15,Latini_PRB15,Wu_PRB15,Qiu_PRB16,Stier_NatComm16,Kidd_PRB16,Kezerashvili_FBS17,Mostaani_PRB17,Szyniszewski_PRB17,Raja_NatComm17,Courtade_PRB17,Donck_PRB17,Stier_PRL18,Filikhin_NanoTec18}, it does not properly model charged excitons (trions). While their calculated binding energies show a strong dependence on whether the ML is encapsulated, supported, or suspended,  the empirical evidence is reversed. The binding energies of trions in MoSe$_2$ and WSe$_2$, for example, have been repeatedly measured in various configurations showing that they are nearly unaffected by the values of $\epsilon_t$ and $\epsilon_b$ (see Table~\ref{tab:exp}). That is, the role of the environment is mitigated.  

We derive a new potential form, taking into account the three atomic sheets that compose an ML-TMD, as illustrated in Fig.~\ref{fig:DielectricEnvironment}(b). Since electrons and holes are restricted to move in the mid-plane of the ML (their wavefunctions are governed by orbitals of the transition-metal atoms), the chalcogen atomic sheets act as a buffer between charged particles in the ML and the outside world. The in-plane polarizability of these buffer layers provide additional screening to the in-plane component of the Coulomb interaction, thereby hindering the electric field from breaking out to the top and bottom layers when $r$ is not much greater than the thickness of the ML, $d$,  or its characteristic screening length. The calculated binding energies of trions strongly depend on the inter-particle interactions when $r \sim d$, resulting in weak dependence on the values of $\epsilon_t$ and $\epsilon_b$. The measured energy separation between the 1s and 2s states of neutral excitons, which represents how similar is the Rydberg series to that of an effective 2D hydrogen model, is also recovered by the new potential. Furthermore, we identify an important difference between the cases of neutral and charged excitons in ML-TMDs. The crystal in the vicinity of a neutral exciton is not distorted since the exciton's Bohr radius is smaller than the polaron radius of electrons or holes in these materials. 
This scenario changes for trions since their nonzero charge distorts the polar crystal in their vicinity. The phonon cloud leads to an increase in the effective mass of the trion, which in turn leads to an increase in their binding energies.  

\section{A phenomenological model for the Coulomb potential in ML-TMDs}

Our approach to the problem is to replace the system shown in Fig.~\ref{fig:DielectricEnvironment}(a) by the one in Fig.~\ref{fig:DielectricEnvironment}(b). The central atomic sheet comprises electron-deficient transition-metal atoms while the top and bottom ones comprises electron-rich chalcogen atoms. In analogy to the treatment of graphene by Cudazzo \textit{et al}.~\cite{Cudazzo_PRB11}, we consider their in-plane polarizabilities, $\chi_{\pm}$. 
The Poisson equation for the bare Coulomb potential follows  \cite{footnote_Poisson}
\begin{eqnarray}\label{Eq:Poissonv2}
\nabla\left[\kappa(z)\nabla\phi(\bm{r}-\bm{r}';z,z')\right] & = &- 4\pi e_1\delta\left(\bm{r}-\bm{r}'\right)\delta\left(z-z'\right)  \nonumber \\ & - & 4\pi \rho_\mathrm{ind}(\bm{r},z)\,,
\end{eqnarray}
where the potential is induced by a point charge ($e_1$), located at $(\bm{r}',z')$, and the relative dielectric constant is
\begin{equation}\label{Eq:DielEnv2}
\kappa(z)=\left\{\begin{array}{ll}
 \epsilon_t & \mathrm{for}\quad z>d/2,\\
 1 & \mathrm{for}\quad -d/2<z<d/2,\\
 \epsilon_b & \mathrm{for}\quad z<-d/2\,.
 \end{array}\right.
\end{equation}
Using the relation $\rho_\mathrm{ind}= \chi_{\pm}\nabla^2_{\bm{r}}\phi$ for the induced-charge density, the 2D Fourier transform of Eq.~(\ref{Eq:Poissonv2}) reads
\begin{eqnarray}\label{Eq:PoissonMomentumv2}
\!\!\!\!&\!&\!\! \frac{\partial}{\partial z}\!\left[\!\kappa(z)\frac{\partial\phi_{\bm{q}}(z,z')}{\partial z}\!\right]\!-\kappa(z)q^2\phi_{\bm{q}}(z,z')=-\frac{4\pi e_1}{A}\delta\left(\!z\!-\!z' \right)  \nonumber \\
\!\!\!\!&\!&\!\! +2 q^2 \! \left[\delta(z)\ell_+ + \delta\!\left(\!z\!-\!\tfrac{d}{4}\right)\!\ell_- + \delta\!\left(\!z\!+\!\tfrac{d}{4}\right)\!\ell_- \! \right]\!\phi_{\bm{q}}(z,z'),
\end{eqnarray}
where $A$ is the area of the ML and $\ell_\pm=2\pi\chi_\pm$. Fixing the point charge to the mid-plane, $z'=0$, one can solve Eq.~(\ref{Eq:PoissonMomentumv2}) with the boundary conditions that $\phi_{\bm{q}}(z,0)$ is continuous and its derivative is piecewise continuous with jumps of $2q^2\ell_+\phi_{\bm{q}}(0,0)-4\pi e_1/A$ at $z=0$ and of $2q^2\ell_- \phi_{\bm{q}}(\pm d/4,0)$ at $z=\pm d/4$. The Coulomb interaction between $e_1$ and  $e_2$  yields 
\begin{equation}\label{Eq:BareCoulomb}
V(q)= e_2 \phi_{\bm{q}}(0,0) = \frac{2\pi e_1e_2}{A\epsilon(q)q},
\end{equation}
where the static dielectric function follows
\begin{equation}\label{Eq:DiFv2}
\epsilon(q)=\frac{1}{2}\left[\frac{N_t(q)}{D_t(q)}+\frac{N_b(q)}{D_b(q)}\right].
\end{equation}
Defining $p_j \equiv (\epsilon_j-1)/(\epsilon_j+1)$ for the top and bottom dielectric constants ($j=b/t$), we get that 
\begin{eqnarray}\label{Eq:DiFv2def}
D_j(q) &=& 1+q\ell_- -q\ell_- (1+p_j)\text{e}^{-\frac{qd}{2}} - (1-q \ell_- ) p_j \text{e}^{-qd}, \nonumber \\
N_j(q) &=& \left(1+q\ell_-\right)\left(1+q\ell_+\right) \nonumber \\ & + & \left[\left(1-p_j\right)-\left(1+p_j\right)q\ell_+\right]q\ell_-\text{e}^{-\frac{qd}{2}} \nonumber \\
    &+& (1-q\ell_-)(1-q\ell_+ )p_j\text{e}^{-qd}.
\end{eqnarray}
The real-space 2D interaction between the two charges in the mid-plane is then found from,
\begin{eqnarray}
V(r) = \frac{A}{4\pi^2} \int \! \! d^2q \, V(q) \text{e}^{i\mathbf{q}\cdot{\mathbf{r}}}  =  e_1e_2 \int_0^{\infty}  \! \! dq \frac{ J_0(qr)}{\epsilon(q)} \,\, , \,\,\,\, \label{eq:2D_potential_Fourier}
\end{eqnarray}
where $J_0$ is the zeroth-order Bessel function. 

The Rytova-Keldysh potential can be recovered when considering the strict 2D limit [$d=0$ in Eq.~(\ref{Eq:DiFv2def})], 
\begin{equation}\label{Eq:eps_K}
\epsilon(q) \, \xrightarrow{ d=0} \, \epsilon_{\text{RK}}(q)=\frac{\epsilon_t + \epsilon_b}{2} + q(\ell_++2\ell_-),
\end{equation}
and upon its insertion in Eq.~(\ref{eq:2D_potential_Fourier}), one recovers the form of $V_{\text{RK}}(r)$ in Eq.~(\ref{eq:Keldysh}) with $r_0=\ell_++2\ell_-$. The use of $\epsilon_{\text{RK}}(q)$ instead of the more rigorous expression in Eq.~(\ref{Eq:DiFv2}) is valid if $d \ll a_B$ where $a_B$ is the effective Bohr radius. When this condition is met, one can consider the range $q \ll 1/d$ in Eq.~(\ref{Eq:DiFv2}) since the exciton wave function in $\mathbf{q}$-space is negligible when $q \gg 1/a_B$. The main drawback of this approximation is that $d\sim~0.6$~nm is only two or three times smaller than $a_B$ in ML-TMDs \cite{Stier_NanoLett17}. Therefore, $V_{\text{RK}}(r)$  is not accurate enough when $r \sim d$, and the correction to the interaction in this range is much needed when studying three or more particle complexes. For example, the binding energy of a trion is measured with respect to that of an exciton plus a faraway third particle (electron or hole). The interaction between the neutral exciton and the third particle at large distances is dipolar in nature, and its relatively fast decay ($\sim$$1/r^2$) implies that  the binding energy of trions and other few-body complexes is governed by inter-particle interactions at short distances.

\begin{figure}
\includegraphics[width=8cm]{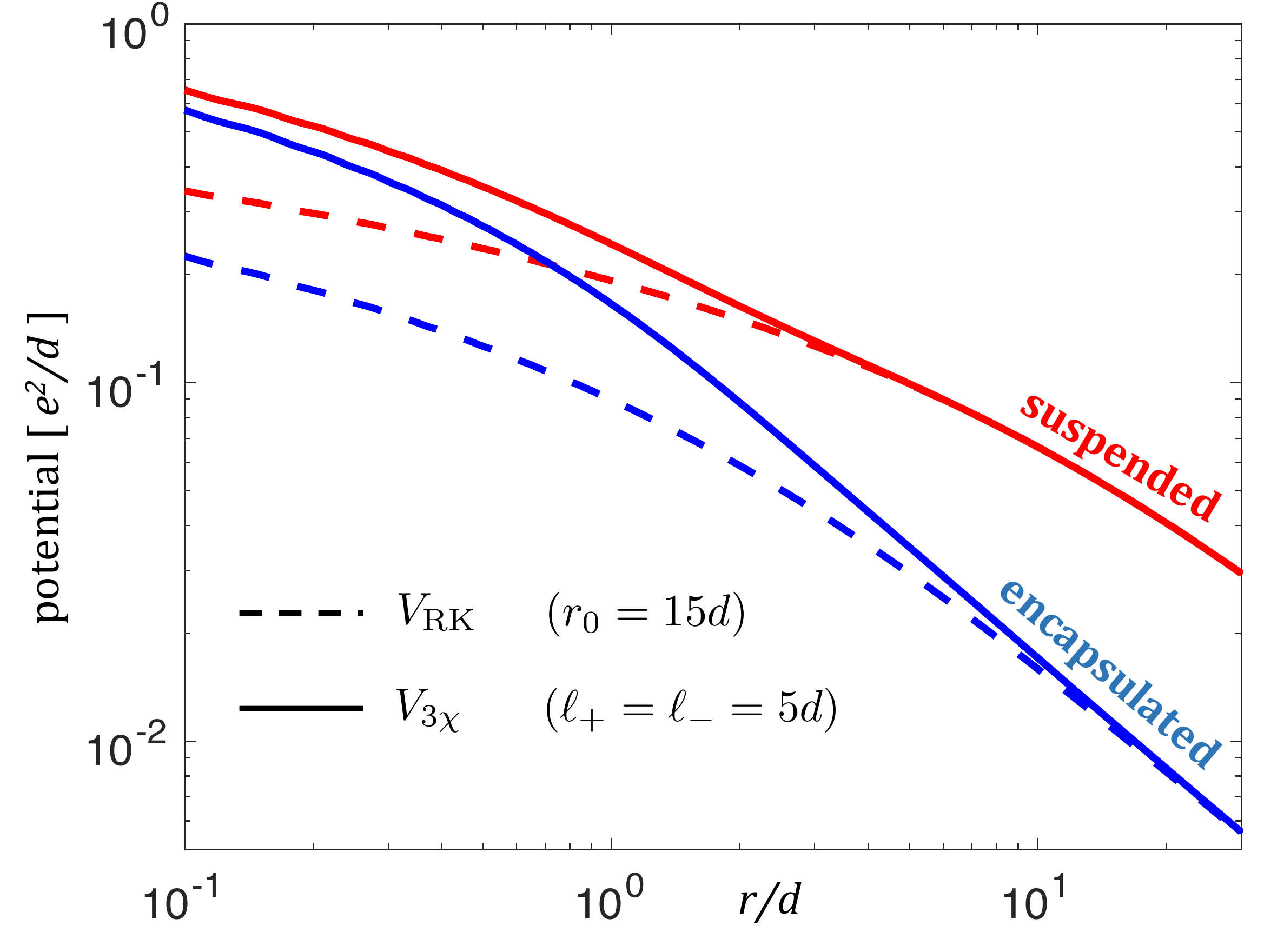}
\caption{The Coulomb potential in suspended ($\epsilon_b=\epsilon_t=1$) and encapsulated ($\epsilon_b=\epsilon_t=6$) monolayers, where $d$\,$=$$\,$0.6~nm. The solid lines show the new potential, $V_{3\chi}(r)$,  when substituting Eq.~(\ref{Eq:DiFv2})~into~(\ref{eq:2D_potential_Fourier}). The dashed lines show the Rytova-Keldysh potential, $V_{\text{RK}}(r)$, provided by Eq.~(\ref{eq:Keldysh}). } \label{fig:potential}
\end{figure}

Figure~\ref{fig:potential}  shows the Rytova-Keldysh potential (dashed lines) and the new potential (solid lines) in two ML configurations. The first one simulates a ML suspended in air, $\epsilon_b=\epsilon_t=1$,  and the second one an encapsulated  ML assuming $\epsilon_b=\epsilon_t=6$.  The results are shown for $r > 0.1d $ since atomic, exchange and correlation effects take over at ultrashort distances \cite{Mostaani_PRB17,Courtade_PRB17}. Hereafter, we denote the new potential by $V_{3\chi}(r)$ owing to the three polarizable atomic sheets of the ML. Two general conclusions can be made by inspection of Fig.~\ref{fig:potential}. Firstly, $V_{\text{RK}}(r)$ and  $V_{3\chi}(r)$ converge to the conventional Coulomb potential, $2e^2/(\epsilon_b + \epsilon_t)r$, when $r \gg d$. Compared with trions, the binding energies of neutral excitons are more affected by the long-range part of the potential because the electron-hole attraction decays relatively slowly ($ \propto 1/r$).  Secondly, the suspended and encapsulated potential forms approach each other faster when $r \lesssim d$ in the case of $V_{3\chi}(r)$ (solid lines) compared with $V_{\text{RK}}(r)$ (dashed lines). This property leads  to a weaker dependence of the trion binding energy on $\epsilon_b$ and $\epsilon_t$ when using $V_{3\chi}(r)$. 

\section{Exciton and trion states in ML-TMDs}

To quantify the effects of $V_{\text{RK}}(r)$ and  $V_{3\chi}(r)$ on excitons and trions, we solve their Schr{\"o}dinger equations.  Within the framework of the effective mass approximation, the Hamiltonian of an $\mathcal{N}$-particle system reads,
\begin{eqnarray}
H_{\mathcal{N}}=  \sum_i^{\mathcal{N}} - \frac{\hbar^2}{2m_i}\nabla^2_i  + \sum_{i<j}^{\mathcal{N}} V(r_{ij})\,,  \label{eq:H}
\end{eqnarray}
where $m_i$ is the effective mass of the $i^{\text{th}}$ particle. The interaction between the $i^{\text{th}}$ and $j^{\text{th}}$ particles, $V(r_{i j})$ where $r_{i j} = | \mathbf{r}_i -  \mathbf{r}_{j}|$,  is modeled by Eq.~(\ref{eq:Keldysh}) when we use $V_{\text{RK}}(r)$, and by substitution of Eq.~(\ref{Eq:DiFv2}) into (\ref{eq:2D_potential_Fourier}) when we use $V_{3\chi}(r)$.  To solve the  Schr{\"o}dinger equation of the $\mathcal{N}$-particle system, we employ the stochastic variational method (SVM), developed by Varga and Suzuki \cite{Varga1995,Varga1998,Varga1997,Varga2008} (see appendix~\ref{sec:SVM} for technical details).  A few important points are in place before we present numerical results.

\subsection{The dielectric constants of hBN and SiO$_2$}

Two of the most common configurations in which excitons are studied in ML-TMDs are either by supporting the ML on SiO$_2$ or by encapsulating the ML in hBN. The dielectric constants of SiO$_2$ are  $\epsilon_{\text{SiO}_2,0} =  3.9$ and $\epsilon_{\text{SiO}_2,\infty} =  2.1$, in the static and high-frequency limits, respectively.  The case of hBN is less clear, where various values for its dielectric constants are reported in the literature. Geick \textit{et al.}  originally reported that the out-of-plane and in-plane dielectric constants with respect to the $c$-axis of bulk hBN are $\epsilon_{\parallel,0}=5.09$ and $\epsilon_{\perp,0}=7.04$ in the low-frequency limit, while the  high-frequency ones are $\epsilon_{\parallel,\infty}=4.1$ and $\epsilon_{\perp,\infty}=4.95$ \cite{Geick_PR66}. Recent first-principles calculations, however, reported that  $\epsilon_{\parallel,0}=3.57$, $\epsilon_{\perp,0}=6.71$,  $\epsilon_{\parallel,\infty}=2.95$ and $\epsilon_{\perp,\infty}=4.87$, showing similar results with those by Geick \textit{et al.} only for $\epsilon_{\perp,0}$ and  $\epsilon_{\perp,\infty}$ \cite{Cai_SSC07}. Dai \textit{et al.} have shown that the first-principles calculated values match very well with their experimental results in which high-quality hBN was used  (see discussion in the supplemental materials of Ref.~\cite{Dai_Science14}). They attributed the differences to the likely presence of misoriented grains in the pyrolytic hBN used by Geick \textit{et al.}  \cite{Geick_PR66}. We follow this explanation and employ the first-principles calculated values because of the high-quality hBN that is used to encapsulate ML-TMDs in recent experiments. The effective values due to the boundary conditions in the Poisson Equation are then $\epsilon_{\text{hBN},0} = \sqrt{\epsilon_{\parallel,0}\cdot \epsilon_{\perp,0}} = 4.9$ and $\epsilon_{\text{hBN},\infty} = \sqrt{\epsilon_{\parallel,\infty}\cdot \epsilon_{\perp,\infty}} = 3.8$. 

Choosing between the static or high-frequency dielectric constants of hBN or SiO$_2$ is  a subtle problem when one wishes to calculate the exciton states in ML-TMDs. The reason is that energies of polar phonons in SiO$_2$ and hBN are of the order of 100~meV \cite{Gunde_PhysB00,Geick_PR66}, lying between the binding energies of the ground and excited states of excitons in ML-TMDs \cite{Stier_PRL18}. The most accurate way to calculate the exciton states in such a scenario is to use a dynamical Bethe-Salpeter Equation of the electron-hole pair function \cite{VanTuan_PRX17}, where the Coulomb potential includes dynamical contributions from polar phonons in the ML and surrounding layers. However, this numerical technique is computationally expensive and cannot be applied to trions without significant approximations. Improving the calculation of the neutral-exciton states because of the polar phonons dynamical effect will be studied in a future work.

As the focus of this work is on trions with emphasis on the  weak dependence of their binding energies on the top and bottom layers, we will employ the high-frequency dielectric constants of hBN and SiO$_2$ in $V_{3\chi}$ or $V_{RK}$. This choice is justified because the polar-phonon energies in hBN and SiO$_2$ are smaller than the trion energy, which is the sum of the relatively large binding energy of the ground-state exciton and the relatively small binding energy of the trion (with respect to the ground-state exciton). Choosing the high-frequency dielectric constants in the trion case is further justified by recalling that the trion is `glued' by short-range forces for which the relative motion of the involved particles is fastest. Accordingly, we assume that atoms in the top and bottom layers cannot track the time-changing electric field lines due to this fast motion.

\begin{table} [b!]
\renewcommand{\arraystretch}{1.25}
\tabcolsep=0.10 cm
\caption{\label{tab:mANDeps}
Effective masses in ML-TMDs \cite{Kormanyos_2DMater15}.}
\begin{tabular}{c|ccccc}
\hline\hline
                         & \text{WSe$_2$}\,\,\,\,\,\,\,\, & \text{MoSe$_2$} \\ \hline
$m_e,\,m_h$ 		& 0.29,\,0.36\,\,\,\,\,\,\,\, & 0.5,\,0.6  \\
$m_{2e},m_{2h}$      & 0.4,\,0.36\,\,\,\,\,\,\,\, & 0.5,\,0.6    \\
\hline \hline
\end{tabular}
\vspace{-2mm}
\end{table}

\subsection{Mass parameters}

So far, we have discussed the screening parameters and dielectric constants that appear in the Coulomb potential. Equally important are the mass parameters because they lead to differences between the binding energies of negative and positive trions in tungsten-based MLs. Table~\ref{tab:mANDeps} lists the effective masses of electrons and holes following ab-initio calculations  \cite{Kormanyos_2DMater15}.  $m_e$ ($m_h$) refers to the electron (hole) mass in a neutral exciton, while $m_{2e}$ ($m_{2h}$) refers to the mass of the added electron (hole)  in a negative (positive) trion. The negative trion in tungsten-based compounds is unique because its electrons have different masses ($m_{e} \neq m_{2e}$ in Tab.~\ref{tab:mANDeps}),  where one electron comes from the top spin-split valley of the conduction band,  while the second electron comes from the bottom one  \cite{Dery_PRB15,Courtade_PRB17}. The masses of the same-charge particles are equal in all other trion cases because they come from time-reversed valleys.  These differences are consequential since the binding energy of the trion is enhanced/suppressed when the added charge is heavier/lighter than the one with the same charge in the neutral  exciton (recall that the trion binding energy is measured with respect to that of the exciton).  In the case of ML-WSe$_2$ , the fact that $m_{2e} > m_e$ while $m_{2h} = m_h$ explains why the measured binding energy of the negative trion is larger than that of the positive one \cite{Courtade_PRB17,Jones_NatPhys16}. It is emphasized that the spin-splitting in the conduction band  is not related to the binding energy of trions  \cite{Courtade_PRB17}.  Appendix~\ref{sec:mass} includes a quantitative analysis of the binding-energy dependence on the effective masses. 

\subsection{The coupling of trions to the lattice}

Additional important mass-related aspect deals with the coupling of trions to the lattice.  The values shown in Tab.~\ref{tab:mANDeps} are the band-edge effective masses, which do not take into account the phonon cloud near a charged particle in polar materials when the atoms move from their equilibrium positions to effectively screen its charge. The phonon cloud increases the effective masses of electrons and holes unless they are bound together in a neutral exciton whose Bohr radius extends over a distance smaller than their polaron radii. The latter are expressed by $r_{\text{e(h)}} \sim \hbar/\sqrt{m_{\text{e(h)}}E_p}$ where $E_p$ is the longitudinal-optical phonon energy. Substituting typical effective mass values (Tab.~\ref{tab:mANDeps}) and phonon energies, $E_p \simeq 30$~meV \cite{Song_PRL13,Dery_PRB15}, the polaron radii are in the range of 2-3 nm in ML-TMDs. The effective Bohr radius of neutral excitons in their ground state is of the order of 1-2 nm \cite{Stier_NanoLett17}, implying that the lattice is largely undistorted in their vicinity due to charge neutrality.  The case of trions is different because of their nonzero charge.  To account for the polaron effect in Eq.~(\ref{eq:H}), we have increased the effective masses of the electrons/holes in a negative/positive trion. This increase leads to a rigid upshift of the trion binding energies for all ML configurations. We will show that very good agreement with experiment is achieved when the effective-mass increase is $\sim$15-25\%. These values are consistent with those found in other chalcogen-based polar semiconductors such as CdS and ZnSe \cite{footnote_mass,Baer_PRA64,Kataria_JAP77,Imanaka_PRB94,Peeters_PRB88}.

\section{Results and comparison with experiment}

Our calculations are focused on ML-MoSe$_2$ and ML-WSe$_2$ for which there are well-established results for the binding energies of trions in various ML configurations  \cite{He_PRL14,Borghardt_PRM17,Courtade_PRB17,Stier_PRL18,Jones_NatPhys16,Branny_APL16,Ross_NatCommun12,Shepard_arXiv17,Wang_NanoLett17,Han_arXiv18}. When benchmarking the calculated values against empirical results, we focus on the trion binding energies and the energy difference between the $1s$ and $2s$ neutral-exciton states, $\Delta_{12}$. The former is directly measured from the energy difference between the spectral lines of the neutral and charged excitons in photoluminescence or reflectivity experiments. Similarly, $\Delta_{12}$ is directly extracted from the energy difference between the spectral lines of the 1s and 2s neutral-exciton states in reflectivity experiments \cite{He_PRL14,Courtade_PRB17,Wang_NanoLett17,Han_arXiv18}. Table~\ref{tab:exp} includes compiled empirical results of WSe$_2$ and MoSe$_2$ in three common ML configurations:  suspended in air, supported by SiO$_2$, and encapsulated in hBN.  Clearly, the trion binding energies show weak dependence on the ML configuration, varying by $\sim$5 meV or less between the different cases.

\begin{table} [t!]
\begin{threeparttable}
\renewcommand{\arraystretch}{1.25}
\tabcolsep=0.18 cm
\caption{\label{tab:exp}
Empirical values of the energy difference between the $1s$ and $2s$ neutral-exciton states ($\Delta_{12}$) and of trion binding energies in ML-WSe$_2$ and ML-MoSe$_2$.  The negative and positive trions are indicated by $X_{\pm}$. The units are in meV. 
While the band structure in tungsten-based MLs supports two types of bound negative trions \cite{Jones_NatPhys16,Plechinger_NatCommun16,Courtade_PRB17,VanTuan_PRX17}, we only list the ground-state binding energy since our model excludes exchange and correlation effects that correspond to their fine structure. }
\vspace{1mm}
\begin{tabular}{r|ccc}
\hline\hline
							& \text{Air}   					&  \text{SiO$_2$}   				&  \text{
							N}  \\ 
							& \text{Suspended}   			& \text{Supported}  				&  \text{Encapsulated}  \\  \hline
\text{WSe$_2$}, $\Delta_{12}$  	&  -              					& $\sim$170\tnote{a} 			&  $\sim$130\tnote{b,c,d} \\
$X_-$ 						& $\sim$39\tnote{*,$\dagger$} 		& 38\tnote{e}        				& 35\tnote{c}  		\\
$X_+$    						& $\sim$26\tnote{$\dagger$}        	& $\sim$23\tnote{e}    			&  21\tnote{c}  	\\ 
\hline \hline
\rule{0pt}{4ex}   
\text{MoSe$_2$}, $\Delta_{12}$  	&  -              					& -							&  $\sim$150\tnote{j} \\
$X_-$ 			&   -       						& $\sim$30\tnote{g}  				&  26\tnote{h}$\,\,,\,\,$30\tnote{i}  	\\
$X_+$ 						&   $\sim$31\tnote{f} 				& $\sim$30\tnote{g}        			& 24\tnote{h}$\,\,,\,\,$30\tnote{i}   \\
\hline \hline
\end{tabular}
\begin{tablenotes}\footnotesize
\item [a] Ref.~[\onlinecite{He_PRL14}], $\qquad$ \item [b] Ref.~[\onlinecite{Borghardt_PRM17}], $\qquad$ \item [c] Ref.~[\onlinecite{Courtade_PRB17}], $\qquad$

\item [d] Ref.~[\onlinecite{Stier_PRL18}], $\qquad$  \item [e] Ref.~[\onlinecite{Jones_NatPhys16}], $\qquad$  \item [f] Ref.~[\onlinecite{Branny_APL16}],$\qquad$

\item [g] Ref.~[\onlinecite{Ross_NatCommun12}],$\qquad$ \item [h] Ref.~[\onlinecite{Shepard_arXiv17}], $\qquad$ \item [i] Ref.~[\onlinecite{Wang_NanoLett17}],$\qquad$

\item [j] Ref.[\onlinecite{Han_arXiv18}],$\qquad \qquad \qquad \qquad$ 
 
\item [*] Private Communication with Xiaodong Xu.$\qquad \qquad \qquad$

\item [$\dagger$] Private Communication with Kin Fai Mak and Jie Shan.
\end{tablenotes}
\end{threeparttable}
\end{table}




\begin{table} [t!]
\renewcommand{\arraystretch}{1.5}
\tabcolsep=0.14 cm
\caption{\label{tab:theory_V3chi}
Calculated values for the same parameters shown in Table~\ref{tab:exp} when using $V_{3\chi}(r)$.  The units are in meV. The first (second) value in the table entries of trions is calculated with (without) the polaron effect. The latter is modeled by a 17\% mass increase of the same-charge particles in the trion complex for ML-WSe$_2$  and 25\% for MoSe$_2$. The fitting parameter in $V_{3\chi}$ is $\ell_{\pm} = 5.9d$ for ML-WSe$_2$ and $\ell_{\pm} = 7.1d$ for ML-MoSe$_2$. } 
\vspace{1mm}
\begin{tabular}{r|ccc}
\hline\hline
& \text{Air}   &  \text{SiO$_2$}   &  \text{hBN}  \\ 
& \text{Suspended}   & \text{Supported}  &  \text{Encapsulated}  \\  \hline
\text{WSe$_2$}, $\Delta_{12}$ &   171.4   &  162.7   &  138.2  \\
$X_-$   & 38.5 (29.4)   & 37.5 (28.5)  & 34.9 (26.1)  \\
$X_+$  & 26.6 (18.7)   & 25.8 (17.9)  & 23.6 (15.9)  \\
\hline \hline
\rule{0pt}{4ex}   
\text{MoSe$_2$}, $\Delta_{12}$  	&  170.3   &  166.7   &  150.9 \\
$X_-$  &   31.5 (18.4) & 31.0 (17.9)  & 29.6 (16.6)  \\
$X_+$ &   29.7 (18.7) & 29.2 (18.2)  & 28.0 (17.0)  \\
\hline \hline
\end{tabular}
\rule{0pt}{0.1ex}   
\end{table}

 







Table~\ref{tab:theory_V3chi} shows the calculated results when using $V_{3\chi}(r)$. A compiled list of all parameters used in these calculations is provided in Appendix~\ref{sec:parameters}.  Each entry for the trion binding energies in the table includes two values in meV where the first (second) one is calculated with (without) the polaron effect. Table~\ref{tab:theory_VRK} shows the respective results when using $V_{\text{RK}}(r)$. The following numerical procedure was used in the calculations. We have first searched for a value of $r_0$ in $V_{\text{RK}}(r)$ or $\ell=\ell_{\pm}$ in $V_{3\chi}(r)$ to best match the empirical results of $\Delta_{12}$. We then use this fitting parameter to calculate the more computationally demanding trion states. The mass increase of the same-charge particles in the trion due to the polaron effect is then used as a second fitting step to match the empirical trion binding energies. Comparing the results in Tabs.~\ref{tab:theory_V3chi} and \ref{tab:theory_VRK}, we find that $V_{3\chi}(r)$ yields better agreement with experiment. Below we discuss a few noticeable features.  

Firstly,  the results in Tabs.~\ref{tab:theory_V3chi} and \ref{tab:theory_VRK} show that the variation in the trion binding energies between suspended, supported and encapsulated is not affected in the second fitting step in which we increase the mass of the same charge particles. That is, the trion binding energies without the mass increase are smaller than the empirical values by $\sim$8~meV in all of the ML-WSe$_2$ configurations and by $\sim$12~meV in all of the ML-MoSe$_2$ configurations. The fact that a mass increase in the ballpark of 15\%-25\% is needed to match the empirical data, reinforces the polar nature of TMDs and is inline with the mass increase that one finds in other chalcogen-based semiconductors \cite{footnote_mass,Baer_PRA64,Kataria_JAP77,Imanaka_PRB94,Peeters_PRB88}. In addition, a larger mass increase in ML-MoSe$_2$ than in ML-WSe$_2$ is needed to reach agreement with experiment (25\% versus 17\%). This fact is consistent with the stronger Fr\"{o}hlich interaction in ML-MoSe$_2$ \cite{Sohier_PRB16}. 

Secondly, the results in Tabs.~\ref{tab:theory_V3chi} and \ref{tab:theory_VRK} show that the calculated values of $\Delta_{12}$ do not perfectly match the experiment results. That is, we could not find a value for $r_0$ in $V_{\text{RK}}(r)$ or $\ell=\ell_{\pm}$ in $V_{3\chi}(r)$ such that $\Delta_{12}\sim130$~meV for ML-WSe$_2$ encapsulated in hBN \cite{Stier_PRL18} and $\Delta_{12}\sim170$~meV when it is supported on SiO$_2$ \cite{He_PRL14}. We attribute this difficulty to the use of high-frequency dielectric constants in hBN and SiO$_2$ in the calculation of the exciton excited states.  Specifically, it is possible that atom vibrations in SiO$_2$ and hBN are fast enough to track the relative motion between the electron and hole in the ML if the exciton is large enough (i.e., in the $2s$ or higher energy states). Indeed, we have reached better agreement when we have simulated a case where the high-frequency dielectric constants, $\epsilon_{\text{hBN},\infty}$ and $\epsilon_{\text{SiO}_2,\infty}$, are used to calculate the  exciton's ground state, while the static-limit values,  $\epsilon_{\text{hBN},0}$ and $\epsilon_{\text{SiO}_2,0}$, are used to calculate its excited states. Using this method and assigning $\ell=6.8d$ in $V_{3\chi}(r)$, we found that $\Delta_{12} = 131.4$~meV for the encapsulated case and $\Delta_{12} = 170.7$~meV for the supported one (Appendix F).

\begin{table} [b!]
\renewcommand{\arraystretch}{1.5}
\tabcolsep=0.14 cm
\caption{\label{tab:theory_VRK}
Calculated values for the same parameters shown in Table~\ref{tab:exp} when using $V_{\text{RK}}(r)$.  The units are in meV. The first (second) value in the table entries of trions is calculated with (without) the polaron effect. The latter is modeled by a 17\% mass increase of the same-charge particles in the trion complex in ML-WSe$_2$  and 25\% in MoSe$_2$.The fitting parameter in $V_{\text{RK}}$ is $r_0= 5.6$~nm for ML-WSe$_2$ and $r_0= 4.9$~nm for ML-MoSe$_2$. } 
\vspace{1mm}
\begin{tabular}{r|ccc}
\hline\hline
& \text{Air}   &  \text{SiO$_2$}   &  \text{hBN}  \\ 
& \text{Suspended}   & \text{Supported}  &  \text{Encapsulated}  \\  \hline
\text{WSe$_2$}, $\Delta_{12}$  	&  215.7   &  187.2   &  114.0 \\
$X_-$  &   43.1 (34.6) & 38.2 (30.3)  & 25.7 (19.7)  \\
$X_+$ &   32.0 (24.5) & 27.9 (20.9)  & 18.0 (12.7)  \\
\hline \hline
\rule{0pt}{4ex}   
\text{MoSe$_2$}, $\Delta_{12}$  	&  258.8   &  229.5   &  149.7 \\
$X_-$  &   44.6 (29.6) & 39.8 (25.7)  & 27.5 (16.3)  \\
$X_+$ &   42.2 (29.6) & 37.7 (25.8)  & 26.0 (16.6)  \\
\hline \hline
\end{tabular}
\rule{0pt}{0.1ex}   
\end{table}

Finally, the trends in Tabs.~\ref{tab:theory_V3chi} and \ref{tab:theory_VRK} are robust. In other words, choosing other parameters for $r_0$ and the mass increase cannot `cure'  the inherent problem of $V_{\text{RK}}(r)$ that we address in this work: A much stronger than observed dependence of the trion binding energies on the ML configuration. The use of $V_{3\chi}(r)$, on the other hand, produces better results. The variation in the trion binding energies between encapsulated and suspended configuration is $\sim$3~meV when using $V_{3\chi}(r)$, while Tab.~\ref{tab:exp}  shows that the empirical variation in of the order of 4-6~meV. As we show in Appendix~\ref{sec:ell}, this slight mismatch can be readily solved by reducing the value of $\ell_-$ while increasing that of $\ell_+$. Here, we chose to use $\ell_+ = \ell_-$ in order to minimize the dependence of the model on fitting parameters. In addition, Appendix~\ref{sec:more_results} includes sets of calculations with different choices of dielectric constants, showing similar trends to the ones in Tabs.~\ref{tab:theory_V3chi}  and ~\ref{tab:theory_VRK}. That is, the dependence of the trion binding energy on the ML configuration is relatively weak (strong)  when using  $V_{3\chi}$ ($V_{\text{RK}}$).

\subsection{Comparison with other models}


While $V_{\text{RK}}(r)$ is the most commonly used potential for calculation of exiton and trion states in ML-TMDs, there are few recent studies that improve this model by considering finite thickness effects with input from ab-initio calculations. For example, Meckbach \textit{et al}. took into account the anisotropy of the effective dielectric constant, finding a quasi-2D Coulomb potential whose Fourier transform in the ML limit reads \cite{Meckbach_PRB18},  
\begin{eqnarray}
V_w(\mathbf{q}) &=&  \frac{2 \pi e^2 \, e^{-qw}}{Aq} \frac{1}{\epsilon_{w}(q)}  \label{Eq:Mechback_Vq}.
\end{eqnarray}
Here, $w$ is a fitting parameter that reflects the wavefunction extension of electrons and holes in the out-of-plane direction. The use of  $e^{-qw}/q$ in the potential can be understood from the 2D Fourier transform of $V(r) \propto 1/ \sqrt{r^2 + w^2}$ when $w \neq 0$. The dielectric function in Eq.~(\ref{Eq:Mechback_Vq}) reads
\begin{eqnarray}
\epsilon_w(q) & = & \frac{ \left( 1 - p_b p_t e^{-2\eta qD}\right) \kappa}{\left( 1 - p_t e^{-\eta qD}\right)\left( 1 - p_b e^{-\eta qD}\right)}  + r_0 q e^{-qw}, \,\,\,\,\,\,\,\,\,\,\,   \label{Eq:Mechback_eps}  
\end{eqnarray}
where $D$ is the nominal thickness of the ML,  $\eta = \sqrt{\epsilon_{\parallel}/\epsilon_{\perp}}$, $\kappa = \sqrt{\epsilon_{\parallel} \epsilon_{\perp}}$ and  $p_{b(t)}  =  (\epsilon_{b(t)}-\kappa)/(\epsilon_{b(t)}+\kappa)$. Here, $\epsilon_{\parallel (\perp)}$ is the effective in-plane (out-of-plane) dielectric constant of the ML.  We have used this potential form in the SVM simulations of the exciton and trion states. The ML parameters for $D$, $\epsilon_{\parallel}$ and $\epsilon_{\perp}$  in $V_w$ are taken from Ref.~\cite{Meckbach_PRB18}  (they are also listed in Appendix~\ref{sec:parameters}), whereas the effective masses in the ML and dielectric constants for SiO$_2$ and hBN are kept as before. We then use both $w$ and $r_0$ as independent fitting parameters to match the empirical results of $\Delta_{12}$, followed by calculation of the trion binding energies with the added polaron effect as a third fitting parameter. The results are shown in Tab.~\ref{tab:theory_Mechback}. Comparing these results with the ones in Tab.~\ref{tab:theory_V3chi}, the agreement with empirical results is better with $V_{3\chi}$, with the additional advantage that fewer free parameters are used. As shown in Appendix~\ref{sec:more_results}, this behavior persists for other choices of the dielectric constants. 


\begin{table} [h]
\renewcommand{\arraystretch}{1.5}
\tabcolsep=0.14 cm
\caption{\label{tab:theory_Mechback}
Calculated values for the same parameters shown in Table~\ref{tab:exp} when using the potential $V_w$. The first (second) value in each pair is calculated with (without) the polaron effect. The latter is modeled by a 17\% mass increase of the same-charge particles in the trion complex of ML-WSe$_2$  and 25\% in MoSe$_2$. The fitting parameters are $r_0 = 4.3$~nm and $w = 2.2$~\AA  ~ for ML-WSe$_2$, and $r_0 = 4$~nm and $w = 0.4$~\AA  ~ for ML-MoSe$_2$.} 
\vspace{1mm}
\begin{tabular}{r|ccc}
\hline\hline
& \text{Air}   &  \text{SiO$_2$}   &  \text{hBN}  \\ 
& \text{Suspended}   & \text{Supported}  &  \text{Encapsulated}  \\  \hline
\text{WSe$_2$}, $\Delta_{12}$  	&  212.2  &  184.7   &  116.4 \\
$X_-$  &   42.2 (34.0) & 37.5 (29.8)  & 26.2  (20.1) \\ 
$X_+$ &   31.5 (24.2) & 27.5 (20.7)  & 18.4  (13.0)  \\ 
\hline \hline
\rule{0pt}{4ex}   
\text{MoSe$_2$}, $\Delta_{12}$  	&  231.7   &  210.1   &  150.4 \\
$X_-$  &   40.0 (26.4) & 36.5 (23.4)  & 27.7  (16.4)  \\
$X_+$ &   37.8 (26.4) & 34.5 (23.5)  & 26.2 (16.6)   \\ 
\hline \hline
\end{tabular}
\rule{0pt}{0.1ex}   
\end{table}

\section{Conclusions}

We have derived a Coulomb potential form that self consistently explains the non-hydrogenic Rydberg series of neutral excitons in monolayer transition-metal dichalcogenides and the weak dependence of the trion binding energies on the dielectric constants of the top and bottom layers.  Its difference from the other potential choices is by considering a phenomenological non-uniform screening profile within the monolayer. We have shown that agreement with experiment is improved by treating the monolayer as three polarizable atomic sheets. When the inter-particle  distance is comparable to the thickness of the monolayer, the chalcogen atomic sheets diminish the effect of the top and bottom dielectric layers on the Coulomb interaction between charged particles in the mid-plane of the monolayer.  

In addition, we have shown the importance of the polaron effect arising from coupling of trions (charged excitons) to the lattice. This effect can be modeled through an increase in the effective mass of charged particles. The mass increase leads to similar enhancement of the trion binding energies in encapsulated, supported, and suspended monolayers, producing an overall far better agreement with the nominal binding energies that one measures in experiment.

To further improve the phenomenological non-uniform screening profile, one should consider the coordination of the chemical bonds in the unit-cell,  leading to a description where both in-plane and out-of-plane components of the screened electric field are affected by the monolayer atomic structure. Further improvements to the model can be achieved by studying the electron-phonon interaction either by its incorporation as part of the few-body Hamiltonian or by evaluating the induced dynamical shift of exciton binding energies. The latter will lead to better quantification of the exciton binding energies in monolayer transition-metal dichalcogenides and their non-hydrogenic nature.

. 


\begin{figure*}[tb!]
\includegraphics[width=16cm]{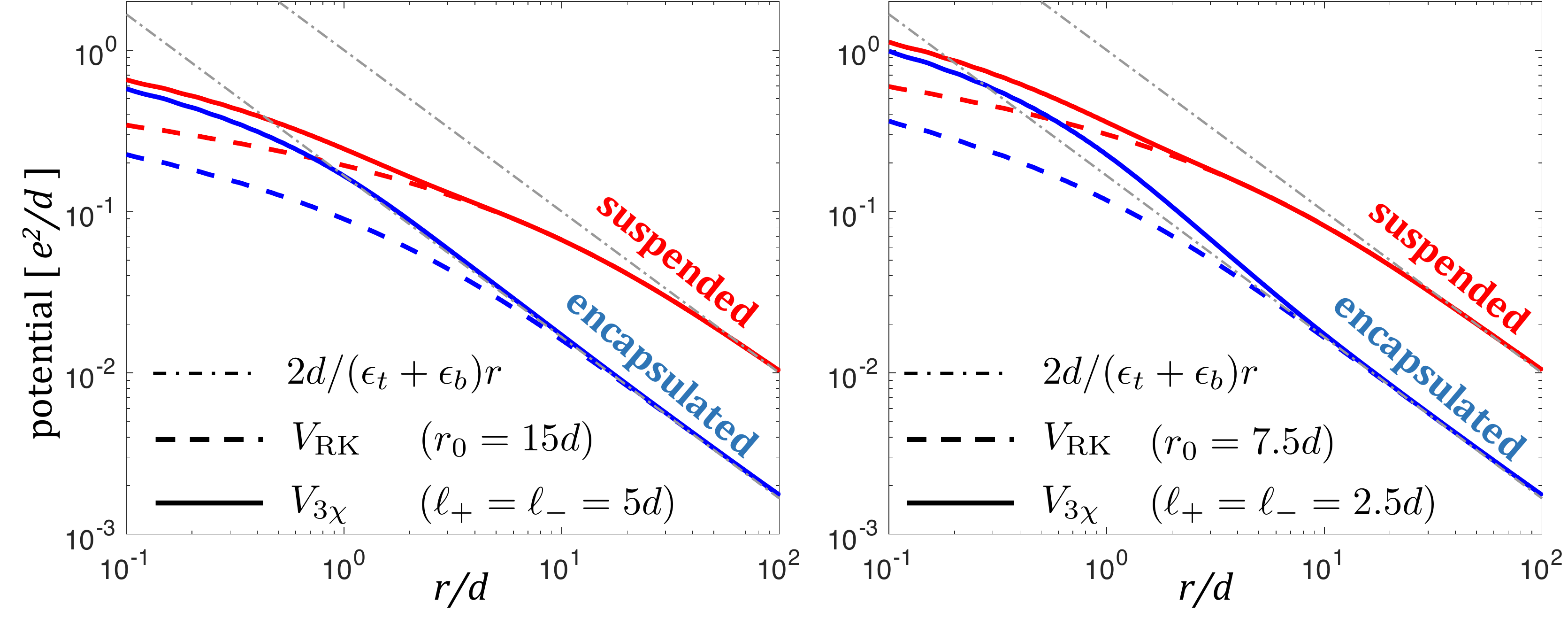}
 \caption{The Coulomb potential in suspended and encapsulated monolayers, modeled by $\epsilon_b=\epsilon_t=1$ and $\epsilon_b=\epsilon_t=6$, respectively. The left panel shows the results when the screening parameters are $\ell_-=\ell_+ = r_0/3 =5d$ and the right panel for $\ell_-=\ell_+ = r_0/3 =2.5d$. The thickness of the monolayer is $d$$\,$=$\,$0.6~nm. The solid lines show the new potential, $V_{3\chi}(r)$, the dashed lines show the Rytova-Keldysh potential, $V_{\text{RK}}(r)$, and the dashed-dotted gray lines show the conventional potential.}  \label{fig:compare_V}
\end{figure*}

\acknowledgments{We are grateful for Xiaodong Xu, Kin Fai Mak, and Jie Shan for sharing photoluminescence results of suspended ML-WSe$_2$ prior to their publication. We also thank Alexey Chernikov and Mikhail Glazov for bringing to our attention the original work of Rytova \cite{Rytova_MSU67}. This work was mostly supported by the Department of Energy under Contract No. DE-SC0014349.  The computational work was also supported by the National Science Foundation (Grant No. DMR-1503601). }


\appendix 


\section{Comparing the new and Keldysh potentials}
Figure~\ref{fig:compare_V} shows the new potential, $V_{3\chi}(r)$, and the Rytova-Keldysh potential, $V_{\text{RK}}(r)$, where the left and right panels show the results for different screening parameters in the monolayer (ML). Regardless of the screening parameters, the two potential forms converge to the conventional Coulomb potential,   $2e^2/(\epsilon_b + \epsilon_t)r$, when $r \gg d$, indicated by the gray dashed-dotted line. The screening parameters of the ML affect the potential mainly at short distances, $r \lesssim d$. In this short-range regime, the suspended and encapsulated forms approach each other, and this trend is stronger for the new potential (the solid lines approach each other faster).

\section{A summary of physical parameters used in the simulations} \label{sec:parameters}
\begin{enumerate}

\item \textit{Effective masses:}

The values of the effective masses in our simulations are listed in Table II of the main text. In Appendix~\ref{sec:mass}, we will provide an analysis of the dependence of the exciton and trion binding energies on the effective mass values, with emphasis on the difference between tungsten and molybdenum based MLs.

\item \textit{Dielectric constants and screening parameters:}

We have used the  high-frequency limit of the dielectric constants in the main text. In Appendix \ref{sec:more_results} we make use of static dielectric constants as well. The values we use in the simulations are 
\begin{enumerate}
\item The static dielectric constant of SiO$_2$ is 3.9, and the high-frequency one is 2.1. 
\item The effective static and high-frequency dielectric constants for hBN are:  $\epsilon_{\text{hBN},0} = \sqrt{\epsilon_{\parallel,0}\cdot \epsilon_{\perp,0}} = 4.9$ and $\epsilon_{\text{hBN},\infty} = \sqrt{\epsilon_{\parallel,\infty}\cdot \epsilon_{\perp,\infty}} = 3.8$.  The values of $\epsilon_{\parallel,0}$, $\epsilon_{\perp,0}$, $\epsilon_{\parallel,\infty}$ and $\epsilon_{\perp,\infty}$ are taken from the work of Cai \textit{et al.} \cite{Cai_SSC07}. To better understand this choice,  see the discussion in the supplemental materials of Ref.~\cite{Dai_Science14}. 

\item When simulating the exciton and trion states with $V_w$, we use $\epsilon_{\parallel}=3.36$ (5.01) and $\epsilon_{\perp}=5.16$ (6.07) in ML-WSe$_2$ (MoSe$_2$), following Ref.~\cite{Meckbach_PRB18}. 

\item The screening parameters of the ML are treated as fitting parameters. We have used $\ell=\ell_{\pm}$ as a single fitting parameter in $V_{3\chi}$, $r_0$ as a single fitting parameter in  $V_{\textit{RK}}$, and $w$ and $r_0$ as two independent fitting parameters in  $V_{w}$. 

\end{enumerate}

\item \textit{Length scales:}
\begin{enumerate}
\item The thickness of the ML in our simulations with $V_{3\chi}$ is $d=0.6$~nm. It is half the out-of-plane lattice constant in bulk TMDs, taking into account the distance between the chalcogen atomic sheets ($\sim$0.3~nm) and the van der Waals gap regions above and below the ML. 

\item When simulating the exciton and trion states with $V_w$, we use $D=6.5$~\AA,  following Ref.~\cite{Meckbach_PRB18}.

\end{enumerate}

\end{enumerate}



\section{SVM} \label{sec:SVM}

The numerical calculations in this paper are performed with the Stochastic Variational Method (SVM)  \cite{Varga1995,Varga1997,Varga1998,Varga2008}, which has been recently applied to study binding energies of exciton complexes in ML-TMDs \cite{Kidd_PRB16,Zhang_NanoLett15}. Another common method is the Quantum Monte Carlo \cite{Mostaani_PRB17,Foulkes_RMP01}, and the two methods  are similar in the sense that both use trial functions to minimize the ground state energy.

The wavefunction in the SVM is expanded in a variational basis which includes correlated Gaussian functions. This variational basis is optimized in a random trial procedure to minimize the ground state energy of a  few-body system whose Schrodinger equation follows
\begin{equation}
\hat{H} \Psi = E \Psi \quad \mbox{with} \quad  \hat{H}=\sum_{i=1}^{N} \frac{{\bf p}^2_i}{2m_i}   + \sum_{i<j}^N V(r_{ij}).
\label{SchrodingerEq.}
\end{equation}
$\{m_i\}$ is the set  of effective masses of the $N$-particle system, and $V(r_{ij})$ is the interaction potential of two particles $i$ and $j$. It is easier to solve the problem by changing from position to Jacobi coordinates, ${\bf x}^T=U {\bf r}^T$, where the transformation matrix reads  \cite{Varga1998,Varga1995}
\begin{equation}
U=
 \begin{pmatrix}
  -1 & 1 & 0 & \cdots & 0 \\
  -\frac{m_1}{\Sigma_{2}} & -\frac{m_2}{\Sigma_{2}} &1  & \cdots & 0 \\
  \vdots  & \vdots  & \vdots & \ddots & \vdots  \\
  -\frac{m_1}{\Sigma_{N-1}} & -\frac{m_2}{\Sigma_{N-1}} &\cdots& \cdots & 1\\
  \frac{m_1}{\Sigma_{N}} & \frac{m_2}{\Sigma_{N}} &\cdots& \cdots & \frac{m_N}{\Sigma_{N}} 
 \end{pmatrix},
\end{equation}
with $\Sigma_{i}= m_1 +m_2 + ...+m_i$. The center of mass coordinate ${\bf x}_N$ is a free degree of freedom if there is no external potential acting on the system. In such a case, the problem  has only $N-1$ variables. The eigenfuntions $\Psi({\bf x})$ are found by the expansion
\begin{equation}
\Psi({\bf x} )=\sum_{i=1}^K c_i \psi({\bf x}, A_i ),
\end{equation}
where the trial basis functions are chosen in the form of correlated Gaussian functions 
\begin{equation}
\psi({\bf x}, A_i )=\mathcal{A} \left\{e^{-\frac{1}{2}{\bf x}{\bf A_i} {\bf x} } \chi \right\}.
\end{equation}
$\chi$  is the spin function and $\mathcal{A}$ is the antisymmetrizer operator. $A_i$ is $(N-1)\times(N-1)$ dimensional symmetric, positive definite matrix whose elements are variational parameters, which will be generated randomly and chosen to optimize the energy level of interest.

The matrix equation corresponding to Eq.~(\ref{SchrodingerEq.}) is
\begin{equation}
HC=EOC,
\end{equation}
where $C=(c_1,c_2, ... , c_K)^T$. The Hamiltonian  and overlap matrix elements are
\begin{eqnarray}
H_{ij} &=& \langle \psi({\bf x}, A_i )\vert \hat{H} \vert \psi({\bf x}, A_j ) \rangle \,\,\,, \nonumber \\
O_{ij} &=& \langle \psi({\bf x}, A_i )\vert \psi({\bf x}, A_j ) \rangle .
\end{eqnarray}
The overlap matrix elements  can be expressed through overlap of correlated Gaussians $G_{A_i}=e^{-\frac{1}{2}{\bf x}{\bf A_i} {\bf x} }$, having the following form in a two-dimensional system
\begin{equation}
\langle G_{A_i}\vert G_{A_j} \rangle=\frac{(2\pi)^{N-1}}{\text{det}(A_i+A_j)}.
\end{equation}
After excluding the center-of-mass term, $P^2/2M$, the kinetic energy can be similarly expressed as
\begin{eqnarray}
&\,& \left\langle G_{A_i}\vert \sum_{i=1}^{N} \frac{{\bf p}^2_i}{2m_i} - \frac{{\bf P}^2}{2M} \vert G_{A_j} \right\rangle=   \langle G_{A_i}\vert G_{A_j} \rangle  \nonumber \,\,\,\,\,\,\,\,\,\,\,\,\\\ 
&\,& \times  \left\{2 \text{Tr}(\Lambda A_i)-2 \text{Tr}\left[(A_i+A_j)^{-1}(A_i\Lambda A_i)  \right] \right\} , \,\,\,\,\,\,\,\,\,\,\,\,\,\,\,
\end{eqnarray}
where $\Lambda$ is an $(N-1)\times(N-1)$ diagonal matrix, $\Lambda_{ij}=\frac{\hbar^2}{2\mu_i}\delta_{ij}$ and $\mu_i=m_{i+1}\Sigma_{i}/\Sigma_{i+1}$  \cite{Varga1997}. The calculation for the potential energy term follows from \cite{Varga1997,Varga2008}
\begin{eqnarray}
\langle G_{A_i}\vert V(r_{\alpha \beta }) \vert G_{A_j} \rangle & \!\!\!=\!\!\! & \int d{\bf r} V({\bf r} )\langle G_{A_i}\vert \delta({\bf r}_\alpha-{\bf r}_\beta-{\bf r}) \vert G_{A_j} \rangle  \nonumber \\ & \!\!\!=\!\!\! & \langle G_{A_i} \vert G_{A_j} \rangle v(c_{\alpha \beta }^{ij}),
\end{eqnarray}
where
 \begin{eqnarray}
\!\! \left( \! c_{\alpha \beta}^{ij} \!\right)^{-1} \!&=&\! \sum_{k,l=1}^{N-1} \left( U^{-1}_{\alpha k}-U^{-1}_{\beta k} \right) (A_i+A_j)_{k l}^{-1} \left( U^{-1}_{\alpha l}-U^{-1}_{\beta l} \right),  \nonumber \\ 
  v(c)&=& \frac{c}{2\pi} \int V(r) e^{-\frac{c}{2}r^2} d{\bf r}.
  \end{eqnarray}
The integral can be rewritten as
 \begin{eqnarray}
 v(c) &=& \frac{c}{2\pi} e_1e_2 \int_0^\infty \frac{dq}{\epsilon(q)} \int J_0(qr)  e^{-\frac{c}{2}r^2} d{\bf r}  \nonumber \\ &=&   e_1e_2 \int_0^\infty \frac{e^{-\frac{q^2}{2 c}}}{\epsilon(q)} dq, 
\end{eqnarray}
and it is calculated numerically in the cases of $V_{3\chi}(r)$ and $V_w(r)$. When using the Rytova-Keldysh potential with  the static dielectric function  $\epsilon(q) \rightarrow \epsilon_{\text{RK}}(q)=\epsilon_{av}(1+qr_0^*)$, where $r_0^* = r_0/\epsilon_{av}$ and $\epsilon_{av}=(\epsilon_t+\epsilon_b)/2$, the integral can be evaluated through the exponential integral and Dawson special function ~\cite{Kidd_PRB16},
 \begin{eqnarray}
\!\!\!\!\! \!\!\!\!\!  \!\!\!\!\! \!\!\!\!\!  v_{\text{RK}}(c) &\!=\!& \frac{ e_1e_2}{\epsilon_{av}} \frac{ 2\sqrt{\pi}  \text{D}\!\left[\frac{1}{ \sqrt{2c} r^*_0}\right] \! - e^{-\frac{1}{2 c {r_0^*}^2}} \text{Ei}\left[\frac{1}{2 c {r_0^*}^2}\right]}{2 r_0^*} \!. 
\end{eqnarray}

\section{Mass dependence} \label{sec:mass}
Using the fact that the masses of electrons and holes in a given ML-TMD are of similar magnitude (while not exactly the same), we provide formulas to estimate the exciton and trion binding energies for a general set of three similar masses $\{m_i,m_j,m_k\}$ where $m_j$ and $m_k$ are masses of charges with the same sign. The analysis will help us to understand the small (large) energy difference between the binding energies of positive and negative trions in molybdenum-based (tungsten-based) TMDs. It is based on linear expansions around reference points calculated for the case that the two (three) particles in an exciton (a trion) have the same mass 
\begin{equation}
E_X(m_i,m_j) \simeq E_X^0(m)\left[1+\beta \left(\frac{m_i+m_j}{2}-m\right)\right],
\label{LinearX}
\end{equation}
\begin{eqnarray}
E_T(m_i,m_j,m_k) \simeq E_T^0(m) \Bigg[ 1&+&   \eta\! \left(\frac{m_j+m_k}{2}-m \right)   \nonumber \\ &+&   \gamma(m_i-m)  \,      \!\Bigg],
\label{LinearT}
\end{eqnarray}
where $E_X^0(m)$ and  $E_T^0(m)$ are exciton and trion energies when the mass of each particle is $m$. We extract the values of $\{\beta, \gamma, \eta\}$  by fitting the numerical results to the above equations, where $m=0.36m_0$ ($m=0.5m_0$) in ML-WSe$_2$ (ML-MoSe$_2$). The results are listed in Table~\ref{abg}, showing the important property that
\begin{equation}
\beta \simeq 2\eta \simeq 2\gamma \,\,.
\label{Relation}
\end{equation}

\begin{table} [h]
\vspace{-4mm}
\renewcommand{\arraystretch}{1.5}
\tabcolsep=0.14 cm
\caption{\label{abg}
The values of $\{\beta, \gamma, \eta\}$ obtained from linear fits to the numerical data. The unit are in $m_0^{-1}$. Here, we use  high-frequency dielectric constants and $V_{3\chi}(r)$ with $\ell=5.9d$ and $\ell_-=7.1d$ in ML-WSe$_2$ and ML-WSe$_2$, respectively.}
\begin{tabular}{r|ccc}
\hline\hline
& \text{Air}   &  \text{SiO$_2$}   &  \text{hBN}  \\ 
& \text{Suspended}   & \text{Supported}  &  \text{Encapsulated}  \\  \hline
\text{WSe$_2$}, $\beta$ &   0.98   &  1.13   &  1.54 \\
$\gamma$    & 0.48   & 0.55  & 0.75  \\
$\eta$  & 0.49   & 0.57  & 0.78  \\
\hline 
\text{MoSe$_2$}, $\beta$ &   0.67   &  0.77   &  1.01 \\
$\gamma$    & 0.33   & 0.38  & 0.50  \\
$\eta$  & 0.34   & 0.38  & 0.51  \\
\hline \hline
\end{tabular}
\rule{0pt}{0.1ex}   
\end{table}

\begin{figure}[ht]
\centering
\includegraphics*[width=9cm]{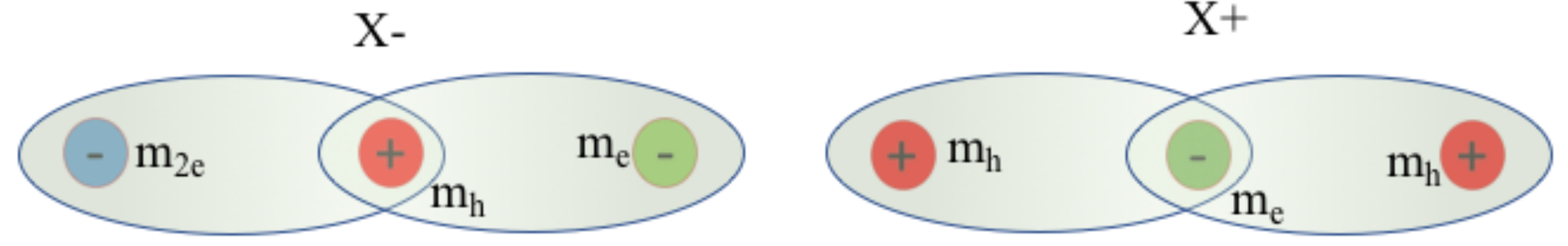}
\caption{Illustration of negative (left) and positive (right) trions in tungsten-based TMDs. Because of the repulsion between charges with the same sign and the resulting larger separation between them, we can view the trion as comprised of two electron-hole pairs, where each contributes about one half to the trion binding energy. The difference in binding energy of the $\{ m_{2e},m_h\}$ pair in $X-$ and the $\{ m_{e},m_h\}$ pair in $X+$ leads to the relatively large difference between the binding energies of negative and positive trions, as written in  Eq.~(\ref{ExplainWSe2}). \label{fig:trion}}
\end{figure}

The trion binding energy is obtained from
\begin{equation}
E_b^{X\pm}=E_X(m_i,m_j)-E_T(m_i,m_j,m_k).
\end{equation}
The difference in the binding energies of positive and negative trions  is
\begin{eqnarray}  \label{DeltaE}
\Delta E_{\pm} &\equiv& E_b^{X+} - E_b^{X-}   \\   &\simeq&  E_T^0(m) \left[ (\eta-\gamma)(m_{h}-m_{e}) -\frac{\eta}{2} \left(m_{2e} -m_{e} \right)\!\right]. \nonumber
\end{eqnarray}
In the case of molybdenum-based TMDs where $m_{2e}=m_{e}$, only the first term contributes. The difference in the binding energies of the positive and negative trions in these compounds is of the order of 1~meV mainly because  $\eta-\gamma$ is a very small quantity. The fact that $m_h$ is larger only by about 20\% than $m_e$ in ML-TMDs also contributes to the small difference. The case of tungsten-based TMDs is different because $m_{2e} \neq m_{e}$, and the second term in Eq.~(\ref{DeltaE}) leads to a large difference such that the binding of the negative trion is larger by $\sim$15~meV. We can also make connection with the exciton binding energy by rewriting Eq.~(\ref{DeltaE}) using the relations in Eqs.~(\ref{LinearX})-(\ref{Relation}),
\begin{eqnarray}
\Delta E_{\pm} & \simeq &  -\frac{\eta}{2} E_T^0(m)  \left(m_{2e} -m_{e} \right)   \simeq  \frac{\beta}{4} E_T^0(m)  \left(m_{e} -m_{2e}\right)    \nonumber \\
& \simeq  & \frac{1}{2} \left[ E_X(m_{e},m_h)-E_X(m_{2e},m_h) \right],
\label{ExplainWSe2}
\end{eqnarray}
where we have used $E_T^0(m) \simeq E_X^0(m) $ in the final step because of the fact that the trion energy does not differ much from the exciton one. $\Delta E_{\pm}$ in tungsten-based TMDs is about half of the gained energy when changing from a lighter exciton $(m_{e},m_h)$ to a heavier one $(m_{2e},m_h)$.  The factor of $\frac{1}{2}$ in Eq.~(\ref{ExplainWSe2})  can be loosely understood as follows: Every opposite-charge pair shares one half of trion energy as illustrated in Fig.~\ref{fig:trion}.

\subsection{The polaron effect} \label{sec:mass_2}
The polaron effect is simulated by a mass increase of charges with the same sign in the trion complex. Following Eq.~(\ref{LinearT}), the binding energy changes by
\begin{eqnarray}
\Delta E_P \simeq \eta  E_T^0(m) \frac{\Delta m_j+\Delta m_k}{2}.
\label{eq:DeltaE_polaron}
\end{eqnarray}
Table \ref{PolaronEffect} lists the change in binding energy, where the first value is the numerically calculated result and the second one (in parentheses) follows Eq.~(\ref{eq:DeltaE_polaron}). The largest deviation between the two values is for $X_+$ in ML-MoSe$_2$ (11 vs $\sim$16~meV), caused by the fact that $\eta$ was evaluated by expansion of the trion energy around $m=0.5m_0$, which better approximates the case of $X_-$, rather than larger values of $m$ which would better fit the case of $X_+$. 

\begin{table} [ht!]
\renewcommand{\arraystretch}{1.5}
\tabcolsep=0.14 cm
\caption{\label{PolaronEffect}
The change in binding energy,  $\Delta E_P$, after increasing the effective mass of the same-charge particles by 17\% in ML-WSe$_2$ and 25\% in ML-MoSe$_2$. The first (second) value is calculated numerically  (extracted from Eq.~(\ref{eq:DeltaE_polaron})).  The high-frequency dielectric constants are used along with $\ell=5.9d$ ($\ell=7.1d$) in $V_{3\chi}(r)$ for ML-WSe$_2$ (ML-WSe$_2$).}
\begin{tabular}{r|ccc}
\hline\hline
& \text{Air}   &  \text{SiO$_2$}   &  \text{hBN}  \\ 
& \text{Suspended}   & \text{Supported}  &  \text{Encapsulated}  \\  \hline
\text{WSe$_2$}, $X_-$   &   -9.1 (-10.0)       &  -9.0 (-10.0)       &  -8.8 (-9.9) \\
                          $X_+$   &   -7.9 (-9.6)     &  -7.9 (-9.6)     &  -7.7 (-9.5)  \\
\hline 
\text{MoSe$_2$}, $X_-$  &   -13.1 (-13.6)   &  -13.1 (-13.5)   & -13.0 (-13.4) \\
                             $X_+$ &   -11.0 (-16.3)   &  -11.0 (-16.2)   & -11.0 (-16.0)  \\
\hline \hline
\end{tabular} 
\end{table}
As one can see, the added binding energy due to the polaron effect is almost independent on the ML configuration. The reason is that the main contribution to the trion binding energy comes from short-range interactions, $r \sim d$, where the environment below and on top of the ML does not play an important role. This behavior is manifested in opposite trends of $\eta$ and $E_T^0(m)$ where the former (latter) is larger when the dielectric constants of the top and bottom layers are relatively large (small). As a result, the product between $\eta$ and $E_T^0(m)$ has a relatively small dependence on the identity of the top and bottom layers. 

\begin{figure}[t!]
\centering
\includegraphics*[width=8.5cm]{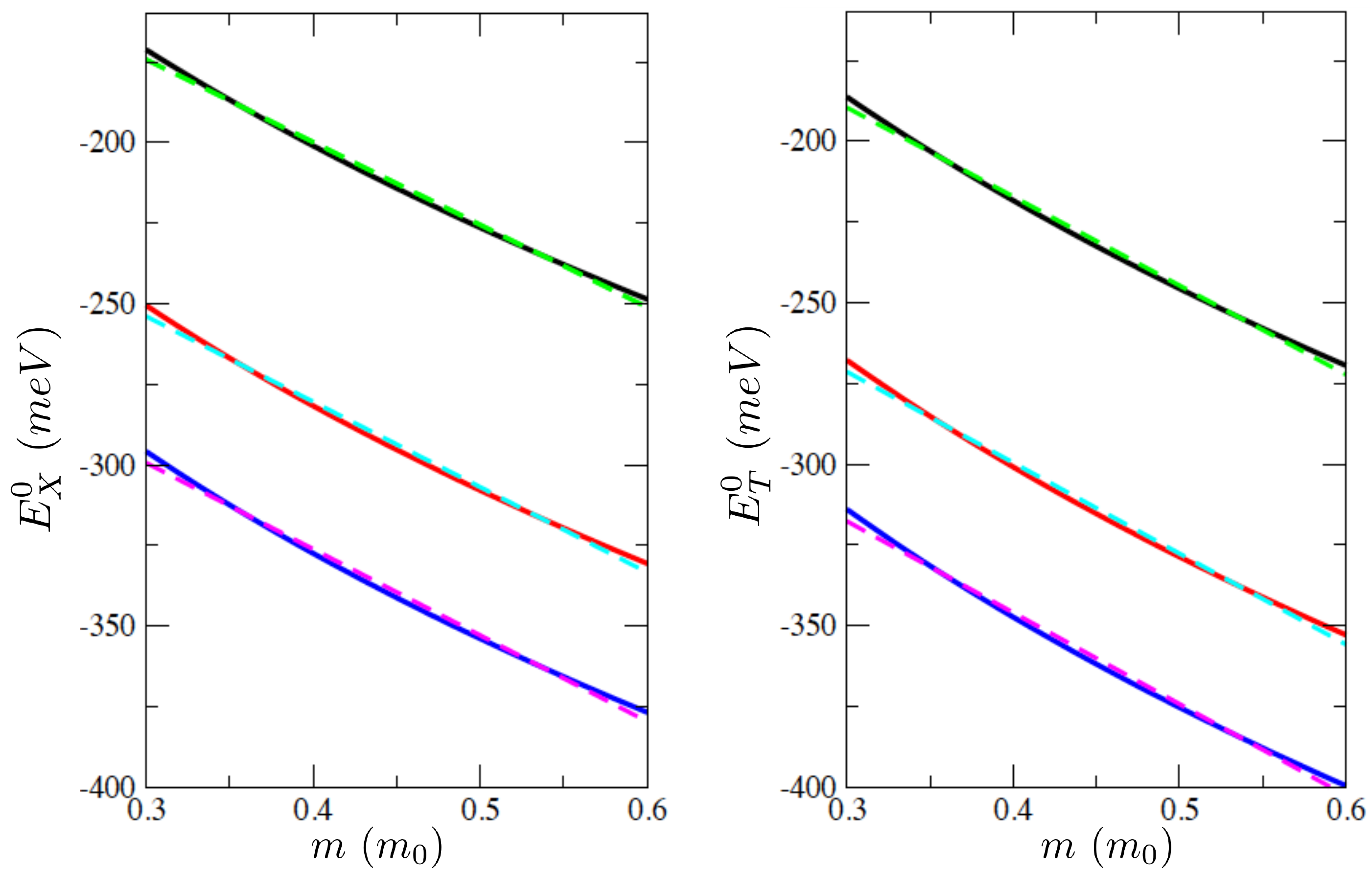}
\caption{$E_X^0(m)$ (left)  and $E_T^0(m)$ (right) as a function of the mass. The bottom/middle/top lines denote the studied cases of suspended/supported/encapsulated MLs. The solid lines denote the numerical data calculated with  $\ell=5.9d$ in $V_{3\chi}(r)$ along with the high-frequency dielectric constants of SiO$_2$ and hBN.  The dashed lines are the linear fits with Eq.~(\ref{LinearXT_m}). }
\label{Fit_m}
\vspace{-4mm}
\end{figure}

\textbf{The values of $E_X^0(m)$ and $E_T^0(m)$}:  The numerical results for the case that the two (three) particles of the exciton (trion) have the same mass can be fitted using a linear approximation, 
\begin{equation}
  E_X^0(m) \simeq A_X^0m +B_X^0\,\,,\,\,\,\,\,\,E_T^0(m) \simeq A_T^0m +B_T^0. \,\,\,\,
\label{LinearXT_m}
\end{equation}
Fig. \ref{Fit_m} shows the linear fits (dashed lines) for the numerical data (solid lines) of exciton (left) and trion (right) energies. 
The error introduced by the linear approximation is less than 4~meV for the range $[0.3m_0,0.6m_0]$.

\section{Screening parameters $l_{\pm}$} \label{sec:ell}
The numerical procedure presented in the main paper was first to fit the empirical values of $\Delta_{12}$, and then use these screening parameters to calculate the binding energies of trions. When using $V_{3\chi}$, we chose to use one fitting parameter, $\ell=\ell_{\pm}$, in the main text to minimize the dependence of the model on fitting parameters. However, if we alleviate the constraint that $\ell_+ = \ell_-$, then there are several sets of parameters that one can use to fit the value of $\Delta_{12}$. Table~\ref{leffect} shows the results for one such representative set: $l_+=7.8d$ and $l_-=3d$ (and using the high-frequency dielectric constants for hBN and SiO$_2$). As one can see from Table~\ref{leffect}, the values of  $\Delta_{12}$ can then match relatively well the experimental values ($\sim$170 and $\sim$130~meV for ML-WSe$_2$ supported on SiO$_2$  and encapsulated in hBN, respectively). However, the trion binding energies calculated with $\{l_+=7.8d, l_-=3d\}$ show increased dependence on the environment, caused by the relatively small value of  $l_-$ (reflecting a smaller screening effect of the chalcogen atomic sheets and therefore a stronger dependence of the trion binding energies on the environment). 
\begin{table}[h]
\renewcommand{\arraystretch}{1.5}
\tabcolsep=0.14 cm
\caption{\label{leffect}
The values of $\Delta_{12}$ and trion binding energies in ML-WSe$_2$ for $l_+=7.8d$ \& $l_-=3d$. The units are in meV. The polaron effect is modeled by a 17\% mass increase of the same-charge particles in the trion complex.} 
\vspace{1mm}
\begin{tabular}{r|ccc}
\hline\hline
& \text{Air}   &  \text{SiO$_2$}   &  \text{hBN}  \\ 
& \text{Suspended}   & \text{Supported}  &  \text{Encapsulated}  \\  \hline
\text{WSe$_2$}, $\Delta_{12}$ &    178.0    &   166.6   &  135.8  \\
$X_-$    &  37.0 (29.1)   &  35.4 (27.6) &   31.6 (24.0) \\   
$X_+$   &  26.6 (19.6)   &  25.2 (18.3) &   21.9 (15.2) \\  
\hline \hline
\end{tabular}
\rule{0pt}{0.1ex}   
\end{table}

\section{Further results with different parameters for the dielectric constants}  \label{sec:more_results}

We show in this appendix that the weak dependence of the trion binding energy on the ML configuration is largely unaffected by the choice of the dielectric constants. Our choice in the main text was to employ the high-frequency dielectric constants of hBN and SiO$_2$ in $V_{3\chi}$, $V_{\text{RK}}$ and $V_w$. 

\subsection{Static dielectric constants for hBN and SiO$_2$}

Tables~\ref{tab:theory_static_3chi}, \ref{tab:theory_static_RK} and \ref{tab:theory_static_Mechback} show the simulated results when choosing to work with the static-limit dielectric constants, $\epsilon_{\text{hBN},0} = \sqrt{\epsilon_{\parallel,0}\cdot \epsilon_{\perp,0}} = 4.9$ and $\epsilon_{\text{SiO}_2,0} =  3.9$,  in $V_{3\chi}$, $V_{\text{RK}}$ and $V_w$, respectively. The first (second) value in the table entries of trions is calculated with (without) the polaron effect. The latter is modeled by a 25\% mass increase of the same-charge particles in the trion complex in MoSe$_2$, and 12.5\% in ML-WSe$_2$, with the exception that the mass increase is 17\% in $V_w$ for ML-WSe$_2$ (this change was needed in order to improve the agreement with experiment).

\begin{table}
\renewcommand{\arraystretch}{1.5}
\tabcolsep=0.14 cm
\caption{\label{tab:theory_static_3chi}
Calculated values when using $V_{3\chi}(r)$ and static-limit dielectric constants for hBN and SiO$_2$.  The units are in meV. The fitting parameter is $\ell_{\pm} = 5.5d$ for ML-WSe$_2$ and $\ell_{\pm} = 6.9d$ for ML-MoSe$_2$.} 
\vspace{1mm}
\begin{tabular}{r|ccc}
\hline\hline
& \text{Air}   &  \text{SiO$_2$}   &  \text{hBN}  \\ 
& \text{Suspended}   & \text{Supported}  &  \text{Encapsulated}  \\  \hline
\text{WSe$_2$}, $\Delta_{12}$ &   182.8    &  162.6   &  139.1  \\
$X_-$    & 38.8 (31.4)   & 36.6 (29.4)  & 34.2 (27.2)  \\
$X_+$   & 26.4 (20.0)   & 24.5 (18.2)  & 22.5 (16.3)  \\
\hline \hline
\rule{0pt}{4ex}   
\text{MoSe$_2$}, $\Delta_{12}$  	&  175.1  &  164.0   &  150.2 \\
$X_-$  &   32.4 (18.9) & 31.3 (17.8)  & 30.0 (16.7)  \\
$X_+$ &   30.5 (19.2) & 29.5 (18.2)  & 28.4 (17.1)  \\
\hline \hline
\end{tabular}
\rule{0pt}{0.1ex}   
\end{table}

\begin{table}
\renewcommand{\arraystretch}{1.5}
\tabcolsep=0.14 cm
\caption{\label{tab:theory_static_RK}
Calculated values when using $V_{\text{RK}}(r)$ and static-limit dielectric constants for hBN and SiO$_2$.  The units are in meV. The fitting parameter is $r_0 =4$~nm for ML-WSe$_2$ and $r_0 = 3.6$~nm for ML-MoSe$_2$.} 
\vspace{1mm}
\begin{tabular}{r|ccc}
\hline\hline
& \text{Air}   &  \text{SiO$_2$}   &  \text{hBN}  \\ 
& \text{Suspended}   & \text{Supported}  &  \text{Encapsulated}  \\  \hline
\text{WSe$_2$}, $\Delta_{12}$ &   287.7    &  192.1   &  110.8  \\
$X_-$    & 55.1 (46.3)   & 39.4 (32.3)  & 25.5 (20.4)  \\    
$X_+$   & 40.2 (32.4)   & 27.6 (21.3)  & 17.1 (12.7)  \\     
\hline \hline
\rule{0pt}{4ex}   
\text{MoSe$_2$}, $\Delta_{12}$  	&  339.4    &  241.2   &  149.6 \\
$X_-$  &   58.6 (38.5) & 49.1 (26.4)  & 29.0 (16.5)  \\ 
$X_+$ &   55.4 (38.5) & 40.7 (26.7)  & 27.4 (16.8)  \\ 
\hline \hline
\end{tabular}
\rule{0pt}{0.1ex}   
\end{table}

\begin{table} [h!]
\renewcommand{\arraystretch}{1.5}
\tabcolsep=0.14 cm
\caption{\label{tab:theory_static_Mechback}
Calculated values when using $V_{w}(r)$ and static-limit dielectric constants for hBN and SiO$_2$.  The units are in meV. The fitting parameters are $r_0 = 3.7$~nm~ and $w = 0.4$~\AA ~ for ML-WSe$_2$, and $r_0 = 3.1$~nm~ and $w = 0.4$~\AA ~ for ML-MoSe$_2$. } 
\vspace{1mm}
\begin{tabular}{r|ccc}
\hline\hline
& \text{Air}   &  \text{SiO$_2$}   &  \text{hBN}  \\ 
& \text{Suspended}   & \text{Supported}  &  \text{Encapsulated}  \\  \hline
\text{WSe$_2$}, $\Delta_{12}$ &   249.1   &  181.2   &  117.4  \\
$X_-$    & 50.1 (40.0)   & 39.2 (30.4)  & 28.9 (21.8)  \\   
$X_+$   & 37.0 (28.1)   & 27.8 (20.1)  & 19.7 (13.5)  \\    
\hline \hline
\rule{0pt}{4ex}   
\text{MoSe$_2$}, $\Delta_{12}$  	&  271.9   &   212.4 &  150.6 \\
$X_-$  &   47.0 (30.8) & 37.9 (23.2)  & 29.2 (16.5)  \\  
$X_+$ &   44.4 (30.8) & 35.8 (23.4)  & 27.6 (16.9)  \\  
\hline \hline
\end{tabular}
\rule{0pt}{0.1ex}   
\end{table}

\subsection{Using mixed dielectric constants in $V_{3\chi}$} \label{sec:other_results} 

The calculated values of $\Delta_{12}$ in the previous simulations did not reach perfect agreement with experiment. Namely, we could not find a value for $r_0$ in $V_{\text{RK}}(r)$ or $\ell=\ell_{\pm}$ in $V_{3\chi}(r)$ such that $\Delta_{12}\sim130$~meV for ML-WSe$_2$ encapsulated in hBN \cite{Stier_PRL18} and $\Delta_{12}\sim170$~meV when it is supported on SiO$_2$ \cite{He_PRL14}. This difficulty can be circumvented by employing the following approach. The exciton ground-state and trion state energies are calculated with the high-frequency dielectric constants, while the exciton excited states are calculated with the static-limit dielectric constants of hBN and SiO$_2$. The motivation for this calculation is that atom vibrations in SiO$_2$ and hBN can be fast enough to track the relative motion between the electron and hole if the exciton is large enough (i.e., in the $2s$ or higher energy states). Table~\ref{tab:theory_static_3chi_low_high} shows the simulated results in ML-WSe$_2$ when using $\ell=6.8d$ in $V_{3\chi}(r)$, where the first (second) value in the table entries of trions is calculated with (without) the polaron effect, modeled by a 25\% mass increase of the same-charge particles in the trion complex. The agreement in this case, for both $\Delta_{12}$ and trion binding energies, is nearly excellent. The relatively small value of $\Delta_{12}$ in the suspended case ($\sim$150~meV) is caused by the relatively large value of the screening parameter in this case ($\ell=6.8d$). The value of $\Delta_{12}$ is larger for the supported case because of the use of different dielectric constants in the calculation of the ground and excited states. 

\begin{table} 
\renewcommand{\arraystretch}{1.5}
\tabcolsep=0.14 cm
\caption{\label{tab:theory_static_3chi_low_high}
Calculated values of ML-WSe$_2$ when using $V_{3\chi}(r)$ with $\ell_{\pm} = 6.8d$. Here, the static-limit dielectric constants for hBN and SiO$_2$ are used in the calculation of the exciton excited states, while the high-frequency ones are used for the trion and ground-state exciton. The units are in meV.} 
\vspace{1mm}
\begin{tabular}{r|ccc}
\hline\hline
& \text{Air}   &  \text{SiO$_2$}   &  \text{hBN}  \\ 
& \text{Suspended}   & \text{Supported}  &  \text{Encapsulated}  \\  \hline
\text{WSe$_2$}, $\Delta_{12}$ &   150.2   &  170.7   &  131.4 \\
$X_-$   & 36.8 (26.1)   &  36.0 (24.9)   & 33.7 (22.9)  \\
$X_+$  & 25.7 (16.4)   &  25.3 (15.7)   & 23.4 (14.0)  \\
\hline \hline
\end{tabular}
\rule{0pt}{0.1ex}   
\end{table}


\begin{table} 
\renewcommand{\arraystretch}{1.5}
\tabcolsep=0.14 cm
\caption{\label{tab:theory_static_Mechback_Ref}
Calculated values of ML-WSe$_2$ when using $V_{w}(r)$ with the all parameters taken from Ref.~\cite{Meckbach_PRB18}, including effective masses,  $m_e=m_h=0.34$, and dielectric constants for hBN and SiO$_2$, $\epsilon_{\text{hBN}} = 2.89$ and $\epsilon_{\text{SiO}_2} =  2.1$. The units are in meV. The best agreement with experiment is reached when the polaron effect is modeled by a 17\% mass increase of the same-charge particles in the trion complex. The fitting parameters are $r_0 = 4.3$~nm and $w=3.6$~\AA.} 
\vspace{1mm}
\begin{tabular}{r|ccc}
\hline\hline
& \text{Air}   &  \text{SiO$_2$}   &  \text{hBN}  \\ 
& \text{Suspended}   & \text{Supported}  &  \text{Encapsulated}  \\  \hline
\text{WSe$_2$}, $\Delta_{12}$ &   202.6   &  174.9   &  128.5 \\
$X_-$   & 40.1 (32.5)   &  35.2 (28.1)   & 27.3 (21.3)  \\
$X_+$  & 30.2 (23.4)   &  26.0 (19.8)   & 19.6 (14.3)  \\
\hline \hline
\end{tabular}
\rule{0pt}{0.1ex}   
\end{table}

\subsection{Other parameter choices for V$_w$} \label{sec:more_other_results} 

We have also simulated the case of ML-WSe$_2$ with $V_w$ using the exact parameters in Ref.~\cite{Meckbach_PRB18}, including the same effective masses and dielectric constants for SiO$_2$ and hBN:  $m_e=m_h=0.34$, $\epsilon_{\text{SiO}_2,\infty} =  2.1$, and $\epsilon_{\text{hBN},\infty} = 2.89$. The latter seems to be the case that $\epsilon_{\text{hBN},\infty} =\epsilon_{\parallel,\infty}$ rather than $\epsilon_{\text{hBN},\infty} = \sqrt{\epsilon_{\parallel,\infty}\cdot \epsilon_{\perp,\infty}}$ \cite{Cai_SSC07,Dai_Science14}, giving rise to a smaller contrast between the suspended and encapsulated configurations. Table~\ref{tab:theory_static_Mechback_Ref} presents the results where the two fitting parameters are $r_0=4.3$~nm and $w=3.6$~\AA, showing that $\Delta_{12}$ is in nearly excellent agreement with experiment. However, despite the seemingly small chosen value for the dielectric constant of hBN, the variation in trion energies is still twice than in the experiment (10-13~meV vs 4-6~meV). 

\subsection{Comparing the potential models}

All in all, comparing the experimental data in Tab.~\ref{tab:exp} with the simulated results, calculated with various parameter and potential choices throughout this work, shows that $V_{3\chi}$ consistently achieves better agreement with the empirical trion binding energies. It requires a single fitting parameter $\ell$ to match the empirical results of a given compound regardless of the ML configuration.





\begin{thebibliography}{99}
\bibitem{footnote_Poisson} The contribution from the nonzero charge of the infinite cationic and anionic sheets to the electric-field vanishes in the mid-plane by symmetry. Thus, we only consider how the polarizabilities of these sheets affect the Coulomb interaction between itinerant charges in the mid-plane.
\bibitem{footnote_mass} The polaron parameters in bulk CdS and ZnSe semiconductors are $\alpha \simeq 0.53$ and $\alpha \simeq 0.43$, respectively \cite{Baer_PRA64,Kataria_JAP77,Imanaka_PRB94}. The correction to the effective mass is then $m^{\ast} \simeq m(1 + \alpha/6)$ in 3D systems and about $m^{\ast} \simeq m(1 + \pi\alpha/8)$  in 2D \cite{Peeters_PRB88}. These values lead to about 20\% increase in the effective masses of electrons and holes in MLs, similar to what we use in this work.
\end{thebibliography}
\end{document}